# Controllable magnetic anisotropy and ferroelasticity in superconducting FeSe monolayer with surface fluorine adsorption

Yueqiao Qu,[1] Yu Liao,[2] Zhixiang Wang,[3] Liang Liu,[4] and Gang Yao[2]*

[1]Key Laboratory of Artificial Structures and Quantum Control (Ministry of Education), School of Physics and Astronomy, Shanghai Jiao Tong University, Shanghai 200240, China
[2]School of Physical Science and Technology, Southwest University, Chongqing 400715, China
[3]Hanhong College, Southwest University, Chongqing 400715, China
[4]School of Physics, State Key Laboratory for Crystal Materials, Shandong University, Jinan 250100, China

*Email: yaogang@swu.edu.cn

**Controllable magnetization in atomically thin two-dimensional magnets is highly desirable for developing spintronics. For FeSe monolayer, its magnetic ground state is not yet fully understood, and the potential in constructing high-speed and advanced devices remains unknown. Using density functional theory calculations, we confirm the spin ordering of monolayer FeSe to be dimer texture. With Fluorine (F) adsorption (F/FeSe), the system exhibits a coverage dependent magnetic anisotropy and multiferroicity which can be attributable to the Jahn-Teller effect, being the benefit to potential spintronic applications. Intriguingly, an inherent coupling between magnetism and ferroelasticity in the most energetically favorable F/FeSe system is proposed. Our study thus not only provides a promising way to control the spintronic properties and construct multiferroics, but also renders F/FeSe an ideal platform for magnetism studies and practical high-performance multifunctional devices.**

With the discovery of the giant magnetoresistance (GMR) effect [1,2], the field of spintronics, which is grounded in spin-polarized charge currents, rapidly emerged as a focus of research. Antiferromagnets, notable for their zero net magnetic moment, offer robust magnetic ordering with long- or quasi-long-range interactions and exceptional resilience to external fields, occupying a pivotal position in both industrial applications and scientific research [3,4]. In these materials, the spin wave, whose polarization depends on the magnetic anisotropy, enables the efficient transport of spin information over long distances within the easy plane or along the preferred easy axis [5,6]. Typically, this spin current can be mediated by strong spin fluctuations or magnons, which are commonly found in antiferromagnetic (AFM) materials [7]. In addition to the famous α-$Fe_2O_3$, long-distance spin transport has also been demonstrated in multiferroic heterostructures integrating ferromagnetism and ferroelasticity [8]. Given these advantageous attributes of antiferromagnets, 2D multiferroic materials exhibiting both antiferromagnetism, with either easy-axis or easy-plane magnetic anisotropy, and ferroelasticity are highly needed but still rare.

Superconductivity, another intriguing property of iron-based materials, has been observed in layered superconductors such as FeSe [9], $A_xFe_{2-y}Se_2$ (A = K, Rb, Cs, and Tl) [10-12], $(Li_{0.8}Fe_{0.2})OHFeSe$ [13], $AFe_2As_2$ (A = Ca, Sr, Ba, and Eu) [14-16], and F-doped XFeAsO (X = Sm, Ce, and La) [17-19]. The tetragonal FeX (X = As and Se) layer, which has been successfully synthesized [20,21], is considered the cornerstone of superconductivity in these materials. Notably, monolayer FeSe exhibits superconductivity with interface-induced high critical temperature ($T_c$) [21,22], demonstrating the potential of this material. Remarkable advancements have been achieved through surface alkali metal adsorption on ultrathin FeSe film, including enhanced $T_c$ [23], superconducting dome [24], exotic phase diagram [25], and the prediction of quantum anomalous Hall effect [26]. Similar to Cu-based superconductors, the magnetic exchange interactions within FeSe layers play a crucial role in superconductivity [27], with the magnetic ground state for FeSe thin films confirmed to be AFM [28]. Given the successful achievement of ultrathin FeSe films through mechanical exfoliation [29] or molecular beam epitaxy [21], and the extensive tunability of this 2D network through various external parameters, it presents an attractive platform not only for studying the interplay between superconductivity and magnetism, but also for unveiling potential applications in spin transport and spintronics devices, yet receiving limited attention. Furthermore, while numerous studies on FeSe nanosheets grown on substrates have been reported, the spin ordering of freestanding FeSe monolayers remains particularly puzzling [27,30-33].

In this work, a comprehensive investigation of the experimental feasibility and magnetic properties of freestanding monolayer FeSe has been conducted using first-principles calculations. The study reveals that this system exhibits antiferromagnetic behavior with a dimer spin texture. Further, the influence of surface adsorption by halogen (F, Cl, Br, and I), alkali metal (Li, Na, K, Ru, and Cs), or alkaline-earth metal (Be, Mg, Ca, Sr, and Ba) atoms, has been systematically explored. Notably, ferroelasticity is observed with F adsorption, we thus mainly discuss the F case in our main text. The magnetic anisotropy of the monolayer can be effectively modulated by the F coverage ($F_c$). Moreover, we explain the in-plane ferroelasticity observed in terms of the Jahn-Teller effect. Particularly, at $F_c$ = 1 monolayer (ML), the magnetization orientation controlled by the ferroelastic (FEL) phase transition is proposed, implying a robust magnetoelastic coupling. Our results presented demonstrate that the potential for the application of 2D FeSe crystal in next-generation memory and computing technologies will be huge.

The density functional theory (DFT) calculations are performed by employing the projector-augmented wave (PAW) potentials [34], as implemented in the Vienna *ab initio* simulation package (VASP) [35]. The generalized-gradient approximation (GGA) of the Perdew-Burke-Ernzerhof (PBE) functional [36] is adopted for the exchange-correlation. The cutoff energy of 450 eV is chosen, and the Brillouin zone is sampled using a 13×13×1 Γ-point centered *k*-mesh. The crystal structures are fully optimized until the forces on each atom are smaller than 0.01 eV/Å, and the difference in total energy between two consecutive steps is <$10^{-6}$ eV. To avoid the interactions between adjacent layers, a vacuum spacing of 20 Å along the *z*-direction is adopted. *Ab initio* molecular dynamic (AIMD) simulations with NVT canonical ensemble are performed within a 3 × 3 × 1 supercell. The total simulation time is 5.0 ps at 1.0 fs per time step. The magnetocrystalline anisotropy (MCA) is evaluated by taking spin-orbit coupling (SOC) into consideration. The charge transfer analysis is conducted using the Bader technique [37].

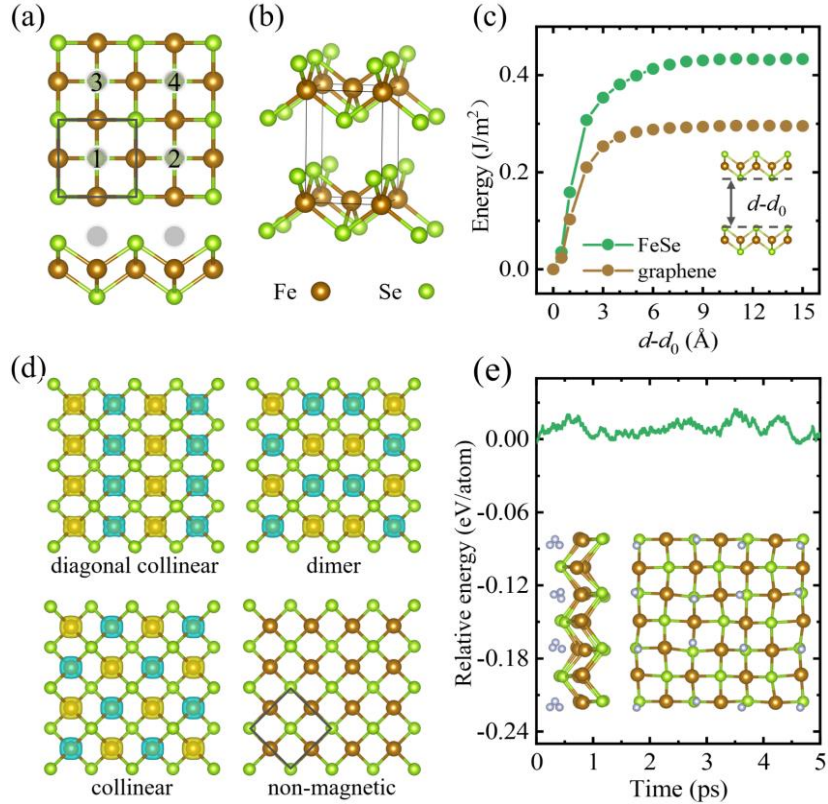

FIG. 1. (a) Crystal structure of monolayer FeSe from top and side views. The unit cell is outlined by the black box. The adsorption sites of the F atom are illustrated by the grey circles labeled by numbers 1 - 4. (b) Crystal structure of bulk FeSe phase. (c) Calculated exfoliation energy of bulk FeSe compared with graphene. Insert: The exfoliation procedure. The $d - d_0$ indicates the separation distance. (d) Three typical AFM magnetic and one non-magnetic configuration on a $2\sqrt{2}\times2\sqrt{2}\times1$ supercell. The colors yellow and blue represent spin up and down, respectively. The black box shows the unit cell. (e) The AIMD simulation based on a 3×3×1 supercell at 300 K. The insert shows the snapshot of the final structure at the end of the simulation.

Figure 1(a) illustrates the crystal structure of monolayer FeSe in different views, showing its $C_{4v}$ point symmetry within the space group of $P4/nmm$ (No. 129). The lattice constants of the optimized structure are $a = b = 3.68$ Å, which is in good agreement with previous experimental reports of FeSe grown on SrTiO$_3$ (FeSe/STO, $a = b = 3.84$ Å [21]). To fabricate the freestanding monolayer FeSe, the most ideal methods are mechanical cleavage and liquid exfoliation from bulk FeSe, where FeSe layers stack along the $c$ axis [Fig. 1(b)]. The exfoliation procedure is modeled in Fig. 1(c), yielding an exfoliation energy ($E_{ex}$) of 0.43 J/m$^2$. Using the same method, the $E_{ex}$ for graphite is calculated to be 0.30 J/m$^2$, which agrees well with experimental measurements (0.32 J/m$^2$) [38], indicating the reliability of the present calculation. Note that ultrathin FeSe films down to bilayer has been successfully achieved very recently by the traditionally mechanical and Al$_2$O$_3$-assisted exfoliation techniques [29], providing the possibility for further research on freestanding monolayer FeSe. As a crucial determinant for the in-plane stiffness of freestanding FeSe monolayer, the 2D Young's modulus is calculated by:

$$Y_{2D} = S_0 \left(\frac{\partial^2 E}{\partial S^2}\right)_{S_0} \quad (1)$$

where $E$ is the energy difference between the strained lattice and the pristine one, $S$ is the strained surface area, and $S_0$ is the pristine surface area. Based on the energy as a function of lattice constant (see Fig. S1 in the Supplemental Material [39]), the $Y_{2D}$ is calculated to be 40.26 N/m. To maintain a stable freestanding structure, a nanosheet should at least be able to withstand its own weight. The out-of-plane deformation $h$ induced by gravity can be calculated by:

$$h/L \approx (\rho g L / Y_{2D})^{1/3} \quad (2)$$

where $Y_{2D}$ is the 2D Young's modulus, $\rho = 3.13 \times 10^{-6}$ kg/m$^2$ is the 2D density of mass of monolayer FeSe, $g$ is the gravitational acceleration (9.81 N/kg), and $L = 100$ μm is the length of the nanosheet. The $h/L$ is calculated to be $4.2 \times 10^{-4}$, comparable to that of the experimentally exfoliated MnPSe$_3$ monolayers ($4.4 \times 10^{-4}$ [48,49]) and graphene ($\sim 3 \times 10^{-4}$ [50]). In summary, the low exfoliation energy, and relatively high in-plane stiffness suggest that the fabrication of freestanding FeSe monolayers is potentially achievable.

Next, to confirm the magnetic ground state, eight different magnetic configurations are considered based on a $2\sqrt{2} \times 2\sqrt{2} \times 1$ supercell: dimer AFM (DAFM), diagonal collinear AFM (DCAFM), collinear AFM (CAFM), bi-collinear AFM (BAFM), chequerboard AFM (CQAFM), diagonal bi-collinear AFM (DBAFM), ferromagnetic (FM), and non-magnetic (NM) [Figs. 1(d) and S5]. Among these, three configurations (DAFM, CAFM, and NM) have previously been proposed as the ground state [27,30-32]. Our calculations reveal that the lowest energy configuration corresponds to the DAFM [Fig. S6(a)], consistent with Refs. [30,33]. Additionally, a detailed discussion of the electronic band structure for this 2D exfoliated nanosheet is provided in the Supplemental Material [39].

Up to now, several fluorinated 2D crystals without signature of clusters or disorders have been reported [41-44], among which tunable ferromagnetic spin ordering has been discovered in MoS$_2$ nanosheet [43] and boron nitride nanotube [44], with $F_c$ as a variable.

We now mainly focus on the FeSe monolayer with F adsorbed on only one side (F/FeSe). The most energetically favorable adsorption site is firstly determined to be site 3 among the six considered ones (Fig. S2), as indicated by the grey circles labeled with numbers 1-4 in Fig. 1(a). Three $F_c$'s (0.25, 0.5, and 1 ML) are selected for the subsequent studies based on a 2×2×1 supercell. Here, the atomic density of the topmost Se-layer is defined as 1 ML. For monolayer FeSe with $F_c$ = 0.5 ML, the energy favorable locations of the two F adatoms are confirmed to be positions 1 and 4. To analyze the stability of F atoms on the surface, the adsorption energies of F/FeSe are calculated as follows:

$$\Delta E = [E(F_x/FeSe) - xE(F) - E(FeSe)]/(x + 8) \tag{3}$$

where $E(F_x/FeSe)$ represents the total energy of $F_x$/FeSe, $E(F)$ and $E(FeSe)$ represent the energies of a single F atom and the FeSe 2×2×1 supercell, respectively. Here, $x$ denotes the F atom number in the supercell. The native $\Delta E$'s summarized in Table I indicate that the adsorption behavior is an exothermic reaction. Importantly, the $\Delta E$ decreases as $F_c$ increases, suggesting that the most stable system is achieved when all the possible positions are occupied (*i.e.*, $F_c$ = 1 ML).

TABLE I. The optimized in-plane lattice constant ($L$), cohesive energy ($E_{coh}$), work function ($\Phi$), and charge transfer ($\Delta Q$) from the Fe atom to the F atom of monolayer FeSe with varying $F_c$.

| $F_c$ (ML) | $L$ (Å) | $E_{coh}$ (eV/atom) | $\Phi$ (eV) | $\Delta Q$ ($e$) |
|---|---|---|---|---|
| 0 | 3.68 | 0 | 4.40 | — |
| 0.25 | 3.71 | -0.14 | 4.96 | 0.07 |
| 0.50 | 3.71 | -0.24 | 5.75 | -0.11 |
| 1.00 | 3.81 | -0.25 | 5.95 | -0.24 |

Given the reactivity of F and the limitations of DFT calculations at zero temperature, concerns naturally arise about whether the FeSe monolayer will be destroyed with F adsorption (e.g., transformation into various iron and selenium fluoride compounds such as $FeF_2$, $FeF_3$, $SeF_4$, $SeF_6$, etc.), or whether disorder and clustering of F adatoms occur regularly in real materials. To address these concerns, the AIMD simulations were performed based on a 3×3×1 supercell at temperatures of 300 K and 500 K [Figs. 1(e) and S3]. The F/FeSe monolayer can maintain its 2D planar networks well with no significant distortion, disorder, cluster, or structural phase transition observed. Also, the system energy changes within a narrow range during the simulation conducted at 300 K. This indicates a high likelihood to obtain this material in the experiment, although the possibility of more substantial reactions or reconstructions cannot be entirely ruled out. Note that the synthesis process of F/FeSe in experiments is more complex and variable than the theoretical prediction, making it challenging for our calculations to simulate the deposition process precisely. Nevertheless, by examining the system energy with varying F-F distances from 1.4 Å, the bond length in $F_2$ gas, to 10.8 Å, the formation of F clusters is energetically unfavorable (Fig. S4), this result further reinforces the possibility of achieving high-quality F/FeSe samples in the experiment.

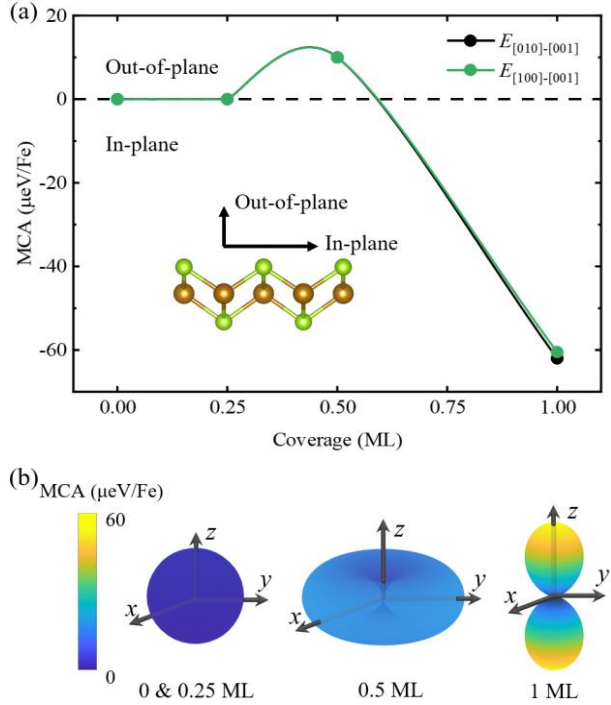

FIG. 2. (a) MCA of F/FeSe as a function of $F_c$. $F_c = 0$ ML indicates the pristine FeSe monolayer. (b) The plots of 3D MCA for F/FeSe with different $F_c$'s.

Fluorine atoms may affect the magnetism because they can manipulate the superconductivity [17,18,51], which competes with antiferromagnetism at low temperatures in some systems [27]. As a key factor for 2D magnets, MCA determines not only the preferred magnetization orientation, but also the strength of the long-range magnetic order against heat fluctuations above zero temperature [52-55]. Due to the magnetic anisotropy, the system energy can be changed by varying the direction of an external magnetic field. Based on this, the MCA can be calculated as MCA = $E_\parallel - E_\perp$, where $E_\parallel$ and $E_\perp$ represent the total energies of the system with in-plane and out-of-plane orientations, respectively. According to the Mermin-Wagner theorem, magnetic ordering with continuous spin symmetries is absent in 2D materials. Thus, 2D magnets usually exhibit a quasi-long-range magnetic ordering rather than a long-range one [54,56]. The former falls under the category of 2D-XY magnets, while the latter belongs to the Ising magnets. In experiments, much attention has been devoted to 2D Ising ferromagnets, such as $CrI_3$ [57] and $Cr_2Ge_2Te_6$ [58], while reports on the XY (anti)ferromagnetic materials are still scarce.

As shown in Fig. 2(a), FeSe monolayer remains an isotropic magnetic character with low $F_c$. As $F_c$ increases, the MCA will increase to around 10 μeV/Fe at $F_c = 0.5$ ML and further decrease to -62 μeV/Fe at $F_c$ = 1 ML, accompanying the transition from the Ising to XY magnet. The preferred magnetization orientations for F/FeSe with varying $F_c$'s are more evident by the 3D plots [Fig. 2(b)]. In contrast, we observed that the DAFM ground state is robust against changes in $F_c$.

A remarkable phenomenon is also observed in our study, which is the F adsorption induced in-plane ferroelasticity. As shown in Fig. 3(a), the energy as a function of elastic strain exhibits two degenerate energy minima at ±8.1% for $F_c$ = 0.5 ML and ±11.8% for 1 ML, being the characteristic of ferroelasticity. Here, the elastic strain refers to the simultaneous application of tensile and compressive strains in the *a* and *b* axis vector directions with equal magnitudes. Similar findings have been reported in hole-doped monolayer XO (X = Sn and Pb) [46,47], which share the same crystal structure with FeSe. For a more direct visualiazation of the FEL phase transition, the unit cells at PA and FEL states are displayed in Fig. 3(b), taking $F_c$ = 1 ML as an example. Furthermore, we investigated the presence of the FEL feature upon adsorption of various elements, including Cl, Br, I, Li, Na, K, Rb, Cs, Be, Mg, Ca, Sr, and Ba. However, no evidence of ferroelasticity is observed (Fig. S11), indicating the importance of proper adatom type in inducing ferroelasticity.

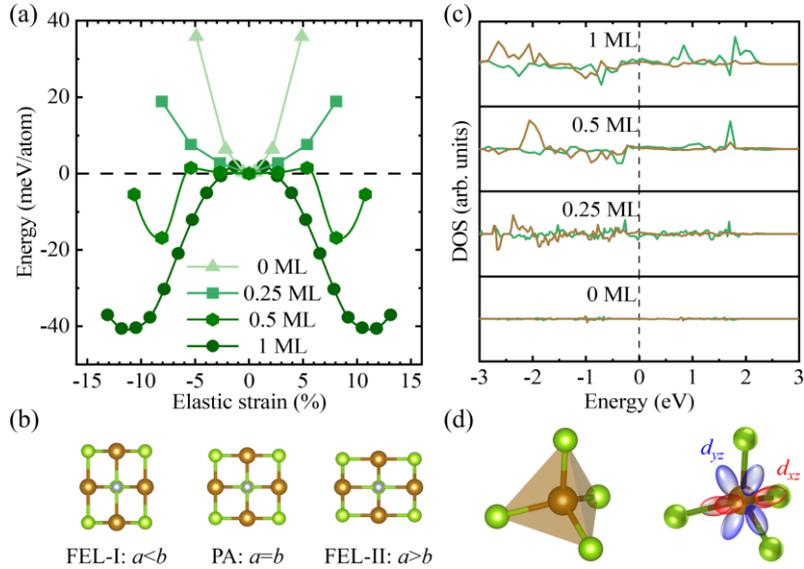

FIG. 3. (a) The energy as a function of the elastic strain for F/FeSe with various $F_c$'s. The energy of the ground state is set to zero. (b) The lattice structures at FEL and PA states. (c) The DOS difference between $d_{xz}$ and $d_{yz}$ orbitals of a single Fe atom with various $F_c$ in F/FeSe at PA state. The spin up and spin down channels are plotted as the color green and brown, respectively. (d) The FeSe$_4$ tetrahedron (left), and the Fe $d_{xz}$ and $d_{yz}$ orbitals (right) in F/FeSe.

Further analysis of the partial density of states (PDOS) of F/FeSe at PA state reveals that the ferroelasticity arises from the Jahn-Teller effect (JTE). The $d_{xz}$ and $d_{yz}$ orbitals of Fe atoms are degenerate in the pristine case ($F_c$=0), while this degeneracy decreases as $F_c$ increases (Fig. S8). The difference between $d_{xz}$ and $d_{yz}$ orbitals with various $F_c$, in Fig. 3(c), can provide a more direct visualization. Clearly, the nearly zero value for $F_c=0$ indicates the high degeneracy, while the non-zero difference after F adsorption signifies the removal of degeneracy. The intensifying fluctuation of this difference with increasing $F_c$ indicates that the degeneracy decreases as $F_c$ increases. As shown in Fig. 3(d), a Fe atom and its four neighboring Se atoms form a FeSe$_4$ tetrahedral structure, with its $d_{xz}$ and $d_{yz}$ orbitals facing the neighboring Se atoms. The degeneracy of these two orbitals in the pristine FeSe monolayer creates a balance of the interactions between Fe atom and its neighboring Se atoms in the two orthogonal directions, maintaining a square lattice. With F adsorption, the removal of this degeneracy disrupts this balance, leading to a tendency for the square lattice to deform into a rectangular shape. This distortion, induced by the removed degeneracy, is known as the JTE [59-61]. Now, a question arises, why does the F/FeSe monolayer maintain a square lattice rather than transform into a rectangular one spontaneously? The answer lies in the two symmetric energy barriers (2.0 meV/atom) near the ground PA state, which keep the energy of the square lattice a local minimum. Furthermore, the square lattice maintains well in the AIMD result at 300 K ($a$=3.82, $b$=3.81 Å), indicating the effectiveness of the barrier. To further validate our understanding of JTE induced ferroelasticity, we also examined the ferroelasticity along the [110] direction using a $2\sqrt{2}\times2\sqrt{2}\times1$ supercell (Fig. S12), and the absence of ferroelasticity along this direction reinforces our conclusion.

Fermi surface nesting induced Peierls distortion constitutes an additional mechanism for ferroelasticity in certain 2D materials, such as MoTe$_2$ [62], WTe$_2$ [63], and GdI$_3$ [64]. In the case of F/FeSe with $F_c$ = 0.5 ML, the observed ferroelasticity, accompanied by the 2×2 structural minimum repetition period (MRP), seems to correspond to the Peierls distortion. However, for $F_c$ values of 0.25 ML and 1 ML, the absence of ferroelasticity and unexpanded MRP, respectively, render this mechanism ineffective. Consequently, the Peierls distortion is excluded as an explanation for our results.

The planar average potential and electron localization function (ELF) of F/FeSe monolayers, presented in Figs. 4(a)-(d), further enrich the JTE understanding. The high and low electron densities for the ELF are indicated by red and blue colors, respectively. Notably, the vacuum region potentials remain flat. The work function ($\Phi$), computed as $\Phi = E_{vacuum} - E_F$, is tabulated in Table I, where $E_{vacuum}$ and $E_F$ denote the vacuum potential and Fermi level of the surface, respectively. Our findings reveal that $\Phi$ increases as the $F_c$ grows, aligning with the conclusion drawn from the adsorption energy $\Delta E$. The gain and loss of electrons for different elements are also illustrated in Fig. 4(e), represented by the positive and negative values, respectively. It is noteworthy that the loss of electrons for Fe atoms ($\Delta Q$) increases with $F_c$ at heavy dosage (Table I), which coincides with the decrease of Fe $d$-orbitals degeneracy. The absence of ferroelasticity in X/FeSe (X = Cl, Br, I, Li, Na, K, Rb, Cs, Be, Mg, Ca, Sr, and Ba) systems may be attributed to insignificant or negligible charge transfer. Furthermore, the $\Delta Q > 0$ for $F_c$ = 0.25 ML also concurs with the absence of ferroelasticity. For direct observation of the charge transfer, the difference charge density ($\Delta\rho$) is shown in Figs. 4(f-h), which defined as $\Delta\rho = \rho_{F/FeSe} - \rho_F - \rho_{FeSe}$, where $\rho_{F/FeSe}$, $\rho_F$, and $\rho_{FeSe}$ represent the charge densities of F/FeSe, monolayer FeSe, and F atoms, respectively.

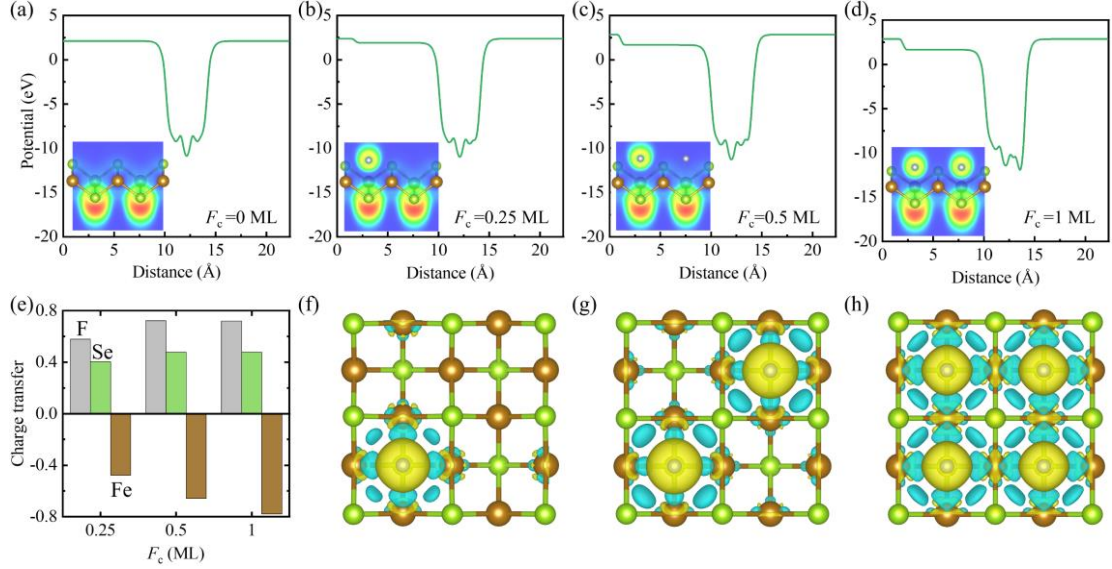

FIG. 4. Planar average potential of F/FeSe monolayer with (a) $F_c = 0$ ML, (b) $F_c = 0.25$ ML, (c) $F_c = 0.5$ ML, and (d) $F_c = 1$ ML. The inserts show the ELF. (e) Charge transfer in F/FeSe. Charge transfer of F/FeSe monolayer with (f) $F_c = 0.25$ ML, (g) $F_c = 0.5$ ML, and (h) $F_c = 1$ ML. The colors yellow and blue represent the gain and loss of electrons, respectively.

For some multiferroics, in-plane magnetic anisotropies are associated with their structural anisotropies [65,66]. Given the ferroelasticity and antiferromagnetism coexist in F/FeSe, we investigate the evolution of MCA under ferroelastic strain with $F_c = 1$ ML (Fig. 5). In the whole path of ferroelastic phase transition, $E_{[001]}$ is consistently higher than $E_{[100]}$ and $E_{[010]}$, indicating an easy plane on the *ab*-plane. Thus, we focus on the energy differences ($E_{[100]\text{-}[010]}$) of $E_{[100]}$ and $E_{[010]}$ for the easy axis evolution. As the lattice transforms into state FEL-I (-II), $E_{[100]\text{-}[010]}$ becomes negative (positive), indicating that the magnetic easy axis is alone the *a* (*b*) axis at state FEL-I (-II). In one word, the elastic strain transforms the intrinsic easy plane into distinct easy axes, with the easy axis always alone the shorter axis vector, delineating an explicit coupling between ferroelasticity and ferromagnetism. Notably, it is reported that the spin transport properties may couple with the easy axis in 2D antiferromagnets [8], this hints at the potential for further studies on spin transport during ferroelastic phase transition.

Previous report has demonstrated successful alkali metal deposition on a single side of freestanding graphene when attached to SiC [67], suggesting that achieving F/FeSe with F adsorbed on a single side might be feasible. Very recently, FeX (X=S, Se, and Te) family with alkali lithium-decorated on both surfaces were predicated to be a stable candidate material for high-temperature quantum anomalous Hall insulators [26]. Note that there exist layered materials composed of FeX and vdW spacer layers, such as LiOH-FeSe. Here, we have also investigated FeSe monolayer with F adsorption on both surfaces. Figures S9 and S13 respectively illustrate the variations of MCA on $F_c$ and elastic energy on strain, which show qualitatively similar results with the case of one side. Based on these results, FeSe monolayer with single-sided and both-sided adsorption of F is intriguing and could stir up a number of further experimental and theoretical studies.

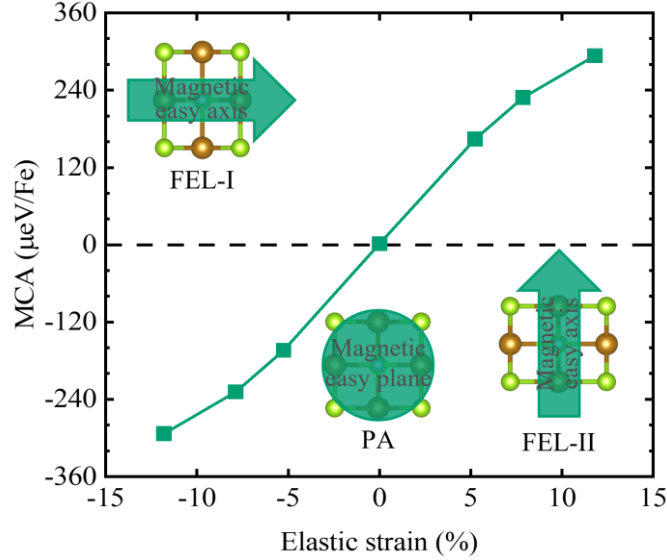

FIG. 5. MCA ($E_{[100]-[010]}=E_{[100]}-E_{[010]}$) as a function of elastic strain for F/FeSe with $F_c$ of 1 ML. Insert: The schematic diagram of the magnetic anisotropy at different states.

In summary, we have systematically investigated monolayer FeSe with F adsorption. First, the pristine monolayer FeSe is confirmed to be an antiferromagnet with dimer spin configuration. Upon F adsorption on the surface, the magnetic anisotropy changes from isotropic character to an out-of-plane easy axis and subsequently to an in-plane easy plane as $F_c$ increases. Meanwhile, the ferroelasticity that emerges in high dosage F/FeSe is attributed to the charge transfer induced Jahn-Teller effect. During the ferroelastic phase transition, the magnetic easy axis is always along the shorter axis vector, indicating the F/FeSe to be multiferroic with strong magnetoelastic coupling. To sum up, we have proposed a promising alternative way to manipulate the spin properties and construct multiferroics with magnetism and ferroelasticity coexisting. Furthermore, these outstanding properties of FeSe monolayer offer valuable opportunities for studying magnetism, superconductivity, or spin transport and applications on spintronic devices.

**REFERRENCE**